\documentclass[12pt,epsf]{iopart}

\usepackage{pstricks}
\usepackage{pst-coil}
\usepackage{amssymb}
\usepackage{graphics,epsf}

\begin{document}

\title{Bounds for scalar waves on self-similar naked-singularity backgrounds.}
\author{Brien C. Nolan}
\address{School of Mathematical Sciences, Dublin City
University, Glasnevin, Dublin 9, Ireland.}
\eads{\mailto{brien.nolan@dcu.ie}}
\begin{abstract}
The stability of naked singularities in self-similar collapse is
probed using scalar waves. It is shown that the multipoles of a
minimally coupled massless scalar field propagating on a spherically
symmetric self-similar background spacetime admitting a naked
singularity maintain finite $L^2$ norm as they impinge on the Cauchy
horizon. It is also shown that each multipole obeys a pointwise
bound at the horizon, as does its locally observed energy density.
$L^2$ and pointwise bounds are also obtained for the multipoles of a
minimally coupled massive scalar wave packet.
\end{abstract}
\submitto{CQG}  \maketitle


\newtheorem{assume}{Assumption}
\newtheorem{theorem}{Theorem}
\newtheorem{prop}{Proposition}
\newtheorem{corr}{Corollary}
\newtheorem{lemma}{Lemma}
\newtheorem{definition}{Definition}
\newtheorem{remarkthe}{Remark}[theorem]
\newtheorem{remarkcorr}{Remark}[corr]
\newtheorem{remarklem}{Remark}[lemma]
\newcommand{\be}{\begin{equation}}
\newcommand{\ee}{\end{equation}}
\newcommand{\so}{{\cal{O}}}
\newcommand{\ch}{{\cal{H}}}
\newcommand{\ce}{{\cal{E}}}
\newcommand{\cf}{{\cal{F}}}
\newcommand{\cm}{{\cal{M}}}
\newcommand{\cmt}{\tilde\cm}
\newcommand{\pnc}{{\cal{N}}}
\newcommand{\bphi}{{\tilde{\phi}}}
\newcommand{\hE}{{\hat{E}}}
\newcommand{\kp}{\kappa}
\def\fin{\hfill \rule{2.5mm}{2.5mm}\\ \vspace{0mm}}
\def\bs{\rule{2.5mm}{2.5mm}}
\section{Introduction}
Spacetimes admitting naked singularities provide test-beds for the
cosmic censorship conjecture. Of course the mere existence of an
example of a spacetime admitting a naked singularity (NS) is not
evidence that the conjecture is invalid, as such spacetimes are
typically unrealistic in that they possess a high degree of
symmetry. Rather, the conjecture may be probed by determining
whether or not these spacetimes (and their NS) are stable under
perturbations. This addresses the question of whether or not NS may
arise from open sets of initial data, which in turn addresses (to an
extent) the question of whether or not NS may arise in nature. For
example, the NS of a charged spherical black hole cannot arise in
nature as it has been shown that the Cauchy horizon accompanying the
singularity is unstable under small perturbations (see \cite{brady}
for a review). However, there is evidence that this instability is
not present for certain spherically symmetric self-similar
spacetimes.

Harada and Maeda have shown the existence of perfect fluid
spacetimes with a soft equation of state that admit naked
singularities but which are stable under perturbations impinging on
the singularity: these have been dubbed the general relativistic
Larson-Penston spacetimes (GRLP) \cite{harada}. (The fate of
perturbations impinging on the Cauchy horizon has not been studied,
and this is an important question.) It has also been shown that
individual massless scalar wave modes of the form $e^{\lambda
v}\phi(x)$ remain finite in the approach to the Cauchy horizon $x\to
x_c$ in a wide class of spherically symmetric self-similar
spacetimes \cite{nolan-waters1}. Here, $v$ is an advanced Bondi
coordinate and $x=v/r$ is a similarity variable (see below; $r$ is
the radius function of the spherically symmetric spacetime). The
same has also been shown to be true for the individual modes of
gravitational (metric and matter) perturbations of the self-similar
Vaidya spacetime \cite{nolan-waters2}.

The aim of the present paper is to expand upon the second of these
three points by studying the propagation of a minimally coupled,
massless scalar field in self-similar spacetimes without recourse to
a mode decomposition (Fourier transform). We find that the results
for individual modes essentially carry over to the multipoles of the
full field, and so have greater significance. (We us the term {\em
multipole} to refer to the coefficients $\phi_{\ell,m}$ of the
spherical harmonics in the standard angular decomposition of a
scalar field $\phi$ on a spherically symmetric background:
$\phi=\sum_{\ell,m}\phi_{\ell,m}Y_\ell^m(\theta,\varphi)$.) The
$L^2$ norm of each multipole field, its pointwise values and its
local energy density remain finite in the approach to the Cauchy
horizon. We consider this to be evidence of the stability of such
naked singularities: the scalar field can be considered to model the
behaviour of gravitational perturbations of the spacetime.

In the following section, we describe the global structure of the
class of spacetimes under consideration: spherically symmetric
spacetimes admitting a naked singularity and whose energy momentum
tensor obeys the weak energy condition. In section 3, we discuss the
$L^2$ and local stability of a massless scalar field propagating in
such a spacetime, and study the local energy density of the field in
section 4. We consider massive fields in section 5, and discuss our
results in section 6. We use the curvature conventions of
\cite{wald}, setting $G=c=1$, and use the notation of \cite{adams}
for Sobolev spaces. A black square $\bs$ indicates the end or
absence of a proof.

\section{Self-similar spherically symmetric spacetimes admitting
a naked singularity.}

We will consider the class of spacetimes which have the following
properties. Spacetime $(\cm,g)$ is spherically symmetric and admits
a homothetic Killing vector field. These symmetries pick out a
scaling origin $\so$ on the central world-line $r=0$ (which we will
refer to as the axis), where $r$ is the radius function of the
spacetime. We assume regularity of the axis to the past of $\so$ and
of the past null cone $\pnc$ of $\so$. We will use advanced Bondi
co-ordinates $(v,r)$ where $v$ labels the past null cones of $r=0$
and is taken to increase into the future. Translation freedom in $v$
allows us to situate the scaling origin at $(v,r)=(0,0)$ and
identifies $v=0$ with $\pnc$. The homothetic Killing field is
\[ {\vec{\xi}}=v\frac{\partial}{\partial v}+r\frac{\partial}{\partial
r}.\] The line element may be written \be ds^2 =
-2Ge^{\psi}dv^2+2e^\psi dvdr+r^2d\Omega^2,\label{lel} \ee where
$d\Omega^2$ is the line element of the unit 2-sphere. The
homothetic symmetry implies that $G(v,r)=G(x), \psi(v,r)=\psi(x)$
where $x=v/r$. The only co-ordinate freedom remaining in
(\ref{lel}) is then $v\to V(v)$; this is removed by taking $v$ to
measure proper time along the regular center $r=0$.

We will not specify the energy-momentum tensor of $(\cm,g)$, but
will demand that it satisfies the null energy condition:
$T_{ab}k^ak^b\geq 0$ for all null vectors $\vec{k}$. Note that this
condition is implied by both the strong energy condition and the
weak energy condition. A complete description of energy conditions
in spherical symmetry is given in \cite{nolan-waters1}. Of these,
the following will be used.
\begin{eqnarray} x\psi^\prime \leq 0,\label{ec1}
\\G^\prime-\psi^\prime(1-xG)G\leq
0.\label{ec2}
\end{eqnarray}


We define the {\em interior region} $\cm_{int}$ of spacetime to be
the interior of $\pnc$, i.e.\ the interior of the causal past of
$\so$. We demand that there are no trapped 2-spheres in $\cm_{int}$.
This is an initial regularity condition, and is equivalent to $G>0$
in $\cm_{int}$. We also note that except for the trivial case of
flat spacetime, there must be a curvature singularity at $\so$
\cite{nolan-waters1}.

The presence of a naked singularity is characterised by the
following result proven in \cite{nolan-waters1}.

\begin{theorem}\label{thmCH}
There exists a future pointing radial null geodesic in $\cm$ with
past endpoint on $\so$ if and only if there exists a positive root
of the equation $xG(x)-1=0$.\fin
\end{theorem}

The features above define the class of spacetimes under
consideration in the remainder of this paper: $(\cm,g)$ is
spherically symmetric and self-similar with line element
(\ref{lel}). There exists a naked singularity, so that there is a
smallest positive root $x_c$ of $xG-1=0$. Then the Cauchy horizon
$\ch$ is $x=x_c$. We take $G,\psi$ to be analytic on
$(-\infty,x_c]$. Einstein's equation and the null energy condition
are assumed, so that (\ref{ec1}) and (\ref{ec2}) apply. The
corresponding conformal diagram is shown in Figure One. There are
subsets of the following categories of spacetimes included in this
class: (i) Vaidya spacetimes (spherically symmetric null dust); (ii)
Einstein-Klein-Gordon spacetimes (spherically symmetric massless
scalar field coupled to gravity); (iii) Lema{\^{\i}}tre-Tolman-Bondi
spacetimes (spherically symmetric dust); (iv) perfect fluid filled
spacetimes (where self-similarity enforces a linear equation of
state - see \cite{Ori-Piran}); (v) Einstein-$SU(2)$ spacetimes
(spherically symmetric $SU(2)$ sigma model coupled to gravity - see
\cite{Bizon-Wasserman}). Of these, the metric functions $G,\psi$ are
explicitly available only for Vaidya spacetime, which has $\psi=0,
2G=1-2\lambda x$, where $\lambda$ is a constant. A naked singularity
occurs for $\lambda\in(0,1/16)$. Among the other cases, the
similarity coordinate $z=t/r$ (where $t$ is proper time along the
fluid flow lines in the dust and perfect fluid cases and a polar
slicing time coordinate in the sigma model case) is more natural
than our $x$, and a nontrivial coordinate transformation is required
to determine $\psi,G$ explicitly. We emphasise however that this
explicit representation is not required for the present purpose.

\begin{figure}[h]
\centerline{\epsfxsize=6cm \epsfbox{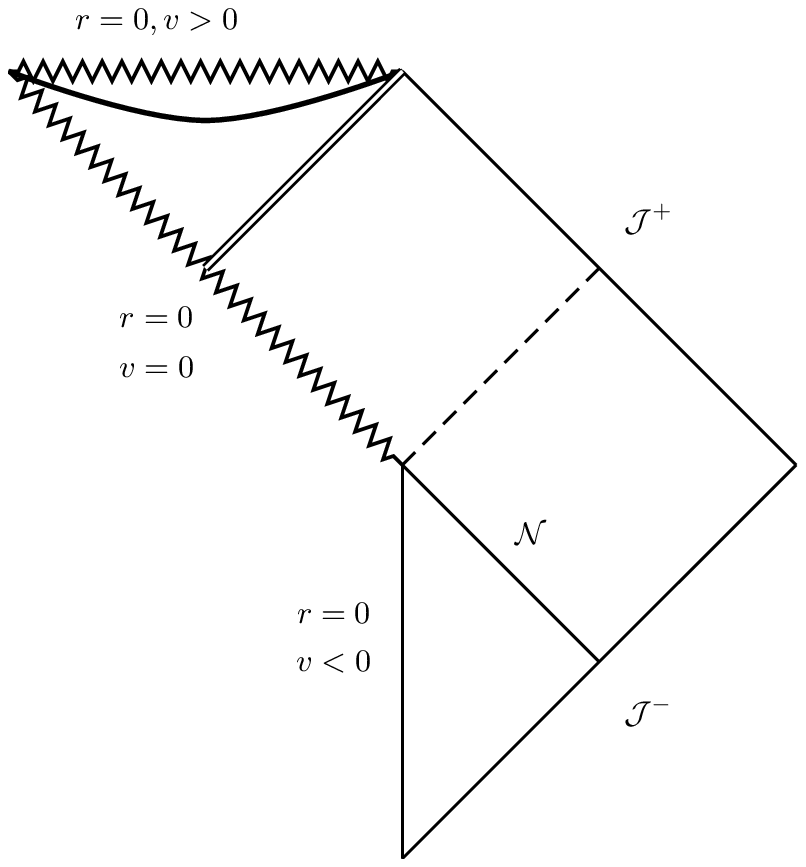}} \caption{Conformal
diagram for an example of a self-similar spacetime admitting a
globally naked singularity. We use the advanced Bondi co-ordinates
$v$ and $r$ described in Section 2. The Cauchy horizon is shown
dashed, the event horizon as a double line and the apparent horizon
as a bold curve. $\pnc$ is the past null cone of the scaling origin.
Other structures can arise; there may be no apparent or event
horizon; the censored portion of the singularity may be null; the
naked portion of the singularity may be time-like.}
\end{figure}

The temporal nature of the surfaces $x=$ constant is central to
the discussion below, and so we note the following which is
immediate from (\ref{lel}).

\begin{lemma}
The surface $x=x_0$ is space-like (respectively time-like, null)
if and only if $x_0(1-x_0G(x_0))>0$ (respectively, $<0,=0$).\fin
\end{lemma}

The definition of $x_c$ then yields the following:

\begin{corr}\label{corr1}
The surfaces $x=$constant are space-like for $x\in(0,x_c)$.\fin
\end{corr}

It is clear that at the past null cone $x=0$ we have $xG-1<0$, and
so a naked singularity forms when the function $G$ meets $1/x$ from
below. We can then show that $1+x_c^2G^\prime(x_c)\geq0$. The case
of equality appears to be quite special: by appealing to the field
equations, it can be ruled out when the energy momentum tensor is
that of dust (Lema\^{\i}tre-Tolman-Bondi spacetime), null dust
(Vaidya spacetime), a massless scalar field or an $SU(2)$ sigma
model. (The case of perfect fluid is more complicated, but by
appealing to a degrees of freedom argument, it appears that equality
is non-generic.) We will therefore assume that equality does not
hold - i.e.\ we will impose the condition \be 1+x_c^2G^\prime(x_c)>0
\label{condition1}\ee as one of the defining characteristics of the
class of spacetimes being studied.

Finally, we give a result from \cite{nolan-waters1} that plays an
important role in the derivation of the stability results below.

\begin{lemma}
$G^\prime< 0$ prior to the formation of a Cauchy horizon and hence
$G^\prime(x_c)\leq 0$.\fin
\end{lemma}

We note that if $G^\prime(x_c)=0$, then
$\left.R_{ab}k^ak^b\right|_{\ch}=0$, where $k^a$ is tangent to the
outgoing radial null direction. This implies that there is no
ingoing radiative flux of energy-momentum crossing the Cauchy
horizon. We rule out this situation as being physically
unrealistic and so we will assume that \be G^\prime(x_c)<0.
\label{condition2}\ee

Note also that this implies that if $G(x_1)=0$ for some $x_1\geq0$,
then $G(x)\leq 0 $ for all $x\geq x_1$ and a naked singularity
cannot arise. Thus we must have $G(x)>0$ in $[0,x_c]$.

\section{$L^2$ and local bounds for the scalar field}

Before proceeding, we make some comments on the analysis that is
carried out below. This analysis is based on the use of naturally
arising space-like slices of the form $x=$ constant, where $x$ is
the homothetic coordinate mentioned above. These surfaces meet the
naked singularity at its past endpoint $(v,r)=(0,0)$ (the scaling
origin). We study the evolution in $x$ of the $L^2-$norm of the
multipoles of a minimally coupled scalar field $\phi$, where the
norm is taken with respect to the measure $d\rho = r^{-1}dr$. (We
deal principally with the massless case, and comment later on the
massive case.) Note then that $\phi|_{x=x_0}\in L^2(\mathbb{R})$
requires vanishing of $\phi$ in the limit as $r\to 0$ on the slice
$x=x_0$. This is an undesirable feature, as this constrains $\phi$
to have support outside the past null cone of the origin. We would
much prefer to be able to deal with the case where $\phi$ is
non-zero on and inside the past null cone. To get around this
problem, we introduce the field $\bphi=\phi e^{\kp\rho}, \kp>0$ and
show that $\bphi$ maintains finite $L^2$ norm in the evolution, at
least for sufficiently small values of $\kp$. This
 allows $\phi$ to be non-zero at $r=0$.

Our strategy runs as follows. By rewriting the wave equation
governing the evolution of the field $\bphi$ in first order
symmetric hyperbolic form, we can apply a standard theorem to
obtain existence and uniqueness on $[x_i,x_c)\times \mathbb{R}$
where $x=x_i\in(0,x_c)$ is an initial slice. Applying another
standard result, we obtain an {\em a priori} bound for a certain
energy norm, but with a coefficient that grows exponentially as
the Cauchy horizon $\ch$ is approached. However, in a
neighbourhood of $\ch$, we can introduce a different energy norm
whose growth is bounded. From this, we prove that $\bphi$ has
finite $H^{1,2}$ norm at the Cauchy horizon, i.e.\ \[
\int_{\mathbb{R}}|\bphi(x_c,\rho)|^2+|\bphi_{,\rho}(x_c,\rho)|^2\,d\rho
<\infty.\] A Sobolev-type inequality then immediately proves
boundedness of $\bphi(x_c,\rho)$ for all $\rho=\ln r\in
\mathbb{R}$. These results apply for all solutions generated by
smooth data with compact support. By taking a sequence of such
solutions, we can generalise to the case where the initial data -
and corresponding solutions - have
$\bphi|_{x={\hbox{constant}}}\in H^{1,2}(\mathbb{R})$. This allows
that $\phi$ is non-zero on the past null cone.

In the remainder of this paper, we use as coordinates $x=v/r$ and
$\rho=\ln r$. The line element then reads
\begin{eqnarray}
ds^2&=&e^{2\rho}[-2Ge^\psi dx^2+2e^\psi(1-2xG)dxd\rho
+2xe^\psi(1-xG)d\rho^2+d\Omega^2].\label{lel2}\end{eqnarray} We
consider the evolution of a massless, minimally coupled scalar field
on the spacetime with this line element, restricting attention to
the region $\cmt=\{\cm:0<x<x_c\}$. Note that by Corollary 1, $x$ is
a time coordinate in this region. We decompose the scalar field
$\phi$ into angular modes: $\phi=\mathrm{Re}{\sum_{\ell,m}\phi_{\ell
m}(x,\rho)Y_{\ell m}(\theta,\varphi)}$. The angular mode indices
$\ell,m$ will be suppressed henceforth, and we take $\phi$ to be the
real part of $\phi_{\ell m}$. Then $\phi$ satisfies the
2-dimensional hyperbolic equation \be
\alpha\phi_{,xx}+2\beta\phi_{,x\rho}+\gamma\phi_{,\rho\rho}+(\alpha^\prime+2\beta)\phi_{,x}+(\beta^\prime+2\gamma)\phi_{,\rho}-\ell(\ell+1)e^\psi\phi=0
\label{weq}\ee and $\bphi=\phi e^{\kp\rho}$ satisfies
\begin{eqnarray}
\alpha\bphi_{,xx}+2\beta\bphi_{,x\rho}+\gamma\bphi_{,\rho\rho}+(\alpha^\prime+2(1-\kp)\beta)\bphi_{,x}+
(\beta^\prime+2(1-\kp)\gamma)\bphi_{,\rho}&&  \nonumber \\
+\left[(\kp^2-2\kp)\gamma-\kp\beta^\prime-\ell(\ell+1)e^\psi\right]\bphi
= 0&& \label{nweq}\end{eqnarray} where
\begin{eqnarray} \alpha&=&\alpha(x)=-2x(1-xG),\label{aldef}\\
\beta&=&\beta(x)=1-2xG,\label{bedef}\\
\gamma&=&\gamma(x)=2G.\label{gadef}
\end{eqnarray}
Here and throughout, a comma denotes a partial derivative. The
fact that the coefficients are independent of $\rho$ is a
consequence of self-similarity. We also note the important fact
that $\alpha<0$ for $x\in(0,x_c)$.

The most notable feature of (\ref{nweq}) - which includes
(\ref{weq}) as the special case $\kp=0$ - is that it is singular
at $\ch$: the space-like surfaces $x=$ constant become
characteristic in the limit $x\to x_c$. This is the fundamental
difficulty in dealing with the equation. In order to obtain an
analytic system, we rescale the time coordinate $x$, placing the
Cauchy horizon at infinity. This introduces the possibility that
the energy will have an infinite amount of time to grow, and so
will diverge at the Cauchy horizon. However, this turns out not to
be the case.

Let $x_i\in(0,x_c)$ and define \be
t=-\int_{x_i}^x\frac{ds}{\alpha(s)}.\label{tdef}\ee By the
definitions above, $t$ is an analytic function of $x$ on
$[x_i,x_c)$, $t(x_i)=0$ and $t\to\infty$ as $x\to x_c^-$. Let \be
{\vec{\varphi}}=\left( \begin{array}{c}
  \bphi \\
  \alpha\bphi_{,x}+\beta\bphi_{,\rho} \\
  \bphi_{,\rho}
\end{array}\right). \label{wdef} \ee
Then (\ref{nweq}) may be written in the first order symmetric
hyperbolic form \be
\vec{\varphi}_{,t}+A\vec{\varphi}_{,\rho}+B\vec{\varphi}=0,\label{system}\ee
where $A,B$ are smooth bounded matrix functions of $t$ on
$(0,\infty)$, and $A$ is symmetric with real, distinct
eigenvalues:
\begin{eqnarray*} A&=& \left(
\begin{array}{ccc}
  0 & 0 & 0 \\
  0 & -\beta & 1 \\
  0 & 1 & -\beta
\end{array}\right),\\
B &=& \left( \begin{array}{ccc}
  0& 1 & -\beta \\
  -\alpha((\kp^2-2\kp)\gamma-\kp\beta^\prime-\ell(\ell+1)e^\psi) & -2(1-\kp)\beta & 2(1-\kp) \\
  0 & 0 & 0
\end{array}\right).\end{eqnarray*}
The following theorem underpins our subsequent work, and is a
standard theorem for symmetric hyperbolic linear systems (see
e.g.\ Chapter 12 of \cite{mcowen}). We recall that for $\Omega$ a
connected subset of $\mathbb{R}^n$, $C^\infty_0(\Omega,X)$ is the
set of smooth functions from $\Omega$ to $X$ which vanish outside
some compact subset of $\Omega$. (When $X=\mathbb{R}$, it will be
omitted.)

\begin{theorem}\label{thmEUforw}
Let $\vec{f},\vec{g}\in C^\infty_0(\mathbb{R},\mathbb{R}^3)$. Then
there exists a unique solution $\vec{\varphi}\in
C^\infty(\mathbb{R}\times[0,\infty),\mathbb{R}^3)$ of the initial
value problem consisting of (\ref{system}) with the initial
conditions
$\vec{\varphi}|_{t=0}=\vec{f},\vec{\varphi}_{,t}|_{t=0}=\vec{g}$.
For each $t>0$, the function
$\vec{\varphi}(t,\cdot):\mathbb{R}\to\mathbb{R}^3$ has compact
support.\fin
\end{theorem}

\begin{corr}\label{corrEUforbphi}
Let $x_i\in(0,x_c)$ and let $f,g\in
C^\infty_0(\mathbb{R},\mathbb{R})$. Then there exists a unique
solution $\bphi\in C^\infty(\mathbb{R}\times[x_i,x_c),\mathbb{R})$
of the initial value problem consisting of (\ref{nweq}) with the
initial conditions $\bphi|_{x=x_i}=f,\bphi_{,x}|_{x=x_i}=g$. For
each fixed $x\in[x_i,x_c)$, the function
$\bphi(x,\cdot):\mathbb{R}\to\mathbb{R}$ has compact support.\fin
\end{corr}

This corollary is worth stating as it provides, in terms of the
natural coordinates, the basic existence and uniqueness theorem
for the field $\bphi$ on the region bounded to the past by the
past null cone and to the future by the Cauchy horizon.

The following corollary provides initial control over the growth
of the $L^2$ norm of $\bphi$, but note that it provides no
information (other than subexponential growth) about the norm in
the limit as the Cauchy horizon is approached. We define \be
\hE(t)=\hE[\vec{\varphi}](t)=\int_\mathbb{R}\|\vec{\varphi}\|^2\,d\rho.\label{hE}\ee
The proof of this corollary is standard, relying on the symmetry
of $A$ and the boundedness of $B$ in (\ref{system}) and the fact
that the solution of Theorem 1 has compact support on each
$t=$constant slice. See for example Theorem 1 of Chapter 12 of
\cite{mcowen}. In this corollary and throughout, the symbols $C_0,
C_1, \dots$ will be used to represent (possibly different)
positive constants that depend only on the metric functions $\psi,
G$ and on the angular mode number $\ell$.

\begin{corr}\label{corrEhat}
Subject to the hypotheses of Theorem \ref{thmEUforw},
$\hE[\vec{\varphi}](t)$ is defined for all $t\geq0$ and satisfies
\[\hE[\vec{\varphi}](t)\leq e^{B_0t}\hE[\vec{\varphi}](0),\]
where $B_0=\sup_{t>0}|-2B|<+\infty$. Consequently,
\begin{eqnarray*}
\int_\mathbb{R}|\bphi(x,\rho)|^2\,d\rho&\leq& e^{B_0t}\hE(0),\\
\int_\mathbb{R}|\bphi_{,\rho}(x,\rho)|^2\,d\rho&\leq&
e^{B_0t}\hE(0),\\
\int_\mathbb{R}|\bphi_{,x}(x,\rho)|^2\,d\rho&\leq&
C_1e^{C_0t}\hE(0).
\end{eqnarray*}\fin
\end{corr}

We note that $B_0$ is finite by virtue of the fact that
$\beta,\psi$ are analytic and bounded on $(0,x_c]$ and that
$\ell,\kappa$ are finite. The bounds on $\bphi$ and its
derivatives come straight from the definition of $\hE(t)$: the
third requires the use of Minkowski's inequality and incorporates
the exponential growth of $\alpha^{-1}$ as $t\to\infty$.

Our next step is to introduce a different energy norm, which can
be shown to display only bounded growth in the approach to the
Cauchy horizon. For this, we return to the original second order
form (\ref{nweq}) of the equation for $\bphi$. For the solutions
of Corollary \ref{corrEUforbphi}, the following integral is
defined and is differentiable for all $x\in[x_i,x_c)$: \be
E(x)=E[\bphi](x)=\int_\mathbb{R}-\alpha\bphi_{,x}^2+\gamma\bphi_{,\rho}^2+
H\bphi^2\,d\rho,\label{E}\ee where $ H =
\ell(\ell+1)e^\psi-(\kp^2-2\kp)\gamma +\kp\beta^\prime.$ From the
definitions of $\beta$ and $\gamma$, we have \[ H =
\ell(\ell+1)e^\psi+2\kp(1-\kp)G-2\kp xG^\prime,\] and so if
$0\leq\kp\leq1$, then $E(x)$ is non-negative for all
$x\in[x_i,x_c)$. The growth of $E(x)$ in the approach to the
Cauchy horizon is controlled by the following result.

\begin{lemma}\label{lemEgrow}
Let
\[0\leq\kp<\min\{1,\displaystyle{\frac{1-x_c^2G^\prime(x_c)}{2}}\}\]
and let
\[m_0>\max\{0,\displaystyle{\frac{1-2\kp}{x_c}},\displaystyle{\frac{-2x_c(\kp
G^\prime(x_c)+x_cG^{\prime\prime}(x_c))}{1-x_c^2G^\prime(x_c)}}
\}.\] Then there exists $x_1\in(x_i,x_c)$ such that $E[\bphi](x)$
is non-negative on $[x_i,x_c)$ and satisfies
\[ \frac{dE}{dx}\leq m_0E\]
for all $x\in(x_1,x_c)$.
\end{lemma}

\noindent{\bf Proof:} Non-negativity of $E$ follows immediately
from its definition and from $\kp\in[0,1)$. Smoothness of $\bphi$
means that $E$ is differentiable, and that its derivative may be
calculated by differentiation under the integral. This is
simplified in three steps: (i) Integration by parts of the term
$\bphi_{,\rho}\bphi_{,\rho x}$ and the removal of a boundary term
- permitted as $\bphi$ has compact support on each slice $x=$
constant. (ii) Removal of the term with $\bphi_{,xx}$ by
application of the equation (\ref{nweq}). (iii) Removal of a total
derivative containing $\bphi_{,x}\bphi_{,\rho x}$. This results in
$\frac{dE}{dx}=\int_\mathbb{R}I\,d\rho$, where
\begin{eqnarray*}
I&=&(\alpha^\prime+4(1-\kp)\beta)\bphi_{,x}^2+2(\beta^\prime+2(1-\kp)\gamma)\bphi_{,x}\bphi_{,\rho}
+\gamma^\prime\bphi_{,\rho}^2\\&&+ [\ell(\ell+1)\psi^\prime
e^\psi-2\kp(\kp
G^\prime+xG^{\prime\prime})]\bphi^2.\end{eqnarray*} For any
$m_0\in\mathbb{R}$, we define $I_R$ by  $I = m_0I_E + I_R$, where
\[
I_E=-\alpha\bphi_{,x}^2+\gamma\bphi_{,\rho}^2+(\ell(\ell+1)e^\psi-(\kp^2-2\kp)\gamma+\kp\beta^\prime)\bphi^2\]
so that $E=\int_\mathbb{R}I_E\,d\rho$. The coefficient of
$\bphi^2$ in $I_R$ is $T(x)= \ell(\ell+1)(\psi^\prime-m_0)e^\psi
-2\kp T_1(x)$, where
\[ T_1=(1-\kp)m_0G+(\kp-m_0x)G^\prime+xG^{\prime\prime}.\] The
assumed bounds on $\kp$ and $m_0$ yield the following:
\begin{eqnarray*}
T_1(x_c) &=&
(1-\kp)\frac{m_0}{x_c}+(\kp-m_0x_c)G^\prime(x_c)+x_cG^{\prime\prime}(x_c)\\
&>&\left(\frac{1-x_c^2G^\prime(x_c)}{2x_c}\right)m_0+\kp
G^\prime(x_c)+x_cG^{\prime\prime}(x_c)>0.
\end{eqnarray*}
Hence by continuity, there exists $x_2\in[x_i,x_c)$ such that
$T_1(x)>0$ for all $x\in(x_2,x_c]$. Then using the energy
condition (\ref{ec1}) and $m_0>0$, we find $T(x)<0$ on $[x_2,x_c]$
and so on this interval,
\[ I_R\leq
(\alpha^\prime+m_0\alpha+4(1-\kp)\beta)\bphi_{,x}^2+2(\beta^\prime+2(1-\kp)\gamma)\bphi_{,x}\bphi_{,\rho}+(\gamma^\prime-m_0\gamma)\bphi_{,\rho}^2.\]
Consider the 1-parameter $(x)$ family of quadratic forms in $R,S$
defined by \begin{eqnarray*}
Q(R,S;x)&=&(\alpha^\prime+m_0\alpha+4(1-\kp)\beta)R^2\\&&+2(\beta^\prime+2(1-\kp)\gamma)RS+(\gamma^\prime-m_0\gamma)S^2.\end{eqnarray*}
The assumption of differentiability of $G(x)$ at the Cauchy
horizon means that the coefficients of $Q(R,S;x)$ are continuous
functions of $x$ at $x=x_c$, and using (\ref{aldef}-\ref{gadef})
we can calculate \begin{eqnarray*}
Q(R,S;x_c)&=&2(2\kp-1+x_c^2G^\prime(x_c))R^2\\&&+\frac{4}{x_c}(1-2\kp-x_c^2G^\prime(x_c))RS+2(G^\prime(x_c)-\frac{m_0}{x_c})S^2.\end{eqnarray*}
The discriminant of this quadratic form is given by
\[
\triangle=\frac{16}{x_c^2}(1-2\kp-x_c^2G^\prime(x_c))(1-2\kp-m_0x_c).\]
Now let $0<\kp<\frac{1-x_c^2G^\prime(x_c)}{2}$ (recall that
$G^\prime(x_c)<0$). If $S=0$, then
\[ Q(R,0;x_c)=2(2\kp-1+x_c^2G^\prime(x_c))R^2\leq0,\]
with equality only if $R=0$. If $S\neq 0$, then the calculation of
the discriminant above shows that $Q(R,S;x_c)$ is negative
definite, provided $1-2\kp-m_0x_c<0$, i.e.\ if $m_0>(1-2\kp)/x_c$.
With these inequalities in place, we see that $Q(R,S;x_c)$ is
negative definite. Hence by continuity, there exists $x_3\in
[x_i,x_c)$ such that $Q(R,S;x)$ is negative definite for all
$x\in[x_1,x_c)$. Hence $I_R\leq 0$ on $[x_1,x_c)$, where
$x_1=\max\{x_2,x_3\}$ giving $I\leq m_0I_E$, and the lemma is
proven.\fin

We can now prove the first of the main results of this section.
\begin{theorem}\label{main1}
Let $\bphi$ be a solution of (\ref{nweq}) that is subject to the
hypotheses of Corollary \ref{corrEUforbphi} and of Lemma
\ref{lemEgrow}. Then the energy $E[\bphi](x)$ of $\bphi$ satisfies
the {\em a priori} bound
\[ E[\bphi](x)\leq C_1\hE[\vec{\varphi}](0),\quad x\in[x_i,x_c].\]  In particular, $\bphi$ has finite
$H^{1,2}$ norm at the Cauchy horizon:
\be\int_\mathbb{R}\bphi^2(x_c,\rho)+\bphi_{,\rho}^2(x_c,\rho)\,d\rho
<C_2\hE[\vec{\varphi}](0)<\infty.\label{hnormch}\ee
\end{theorem}

\noindent{\bf Proof:} From Corollary \ref{corrEhat}, we obtain
\[ E[\bphi](x)\leq c(x)\hE(0),\]
where $c$ is a smooth positive function on $[x_i,x_c)$ that
diverges in the limit $x\to x_c$. Then $0<c(x_1)<\infty$, where
$x_1$ is the value of $x$ identified in Lemma \ref{lemEgrow}. For
$x\in[x_1,x_c)$, we can integrate the result of this lemma to
obtain
\[ E(x)\leq e^{m_0(x-x_1)}E(x_1).\] Combining these two inequalities
yields
\[ E[\bphi](x)\leq C\hE[\vec{\varphi}](0),\quad x\in[x_i,x_c),\]
where $C = e^{m_0(x_c-x_1)}c(x_1)$. Since the right hand side of
the last inequality is independent of $x$, we can take the limit
$x\to x_c$ on the left to see that it applies for all
$x\in[x_i,x_c]$. The inequality (\ref{hnormch}) then follows from
the definition of $E[\bphi](x)$.\fin

To obtain a pointwise bound for $\bphi|_\ch$, we recall the
following Sobolev inequality (see p.\ 1057 of \cite{wald2}).

\begin{lemma}\label{lemWald}
Let $u\in C^\infty_0(\mathbb{R})$. Then for all $s\in\mathbb{R}$,
\be |u(s)|^2\leq\frac12\left\{ \int_\mathbb{R}|u(t)|^2\,dt
+\int_\mathbb{R}|u'(t)|^2\,dt\right\}.\label{embed}\ee\fin
\end{lemma}

\begin{theorem}\label{main2} Let $\bphi$ be a solution of (\ref{nweq}) that is subject to the
hypotheses of Corollary \ref{corrEUforbphi} and of Lemma
\ref{lemEgrow}. Then $\bphi$ is bounded on $[x_i,x_c]\times
\mathbb{R}$.
\end{theorem}

\noindent{\bf Proof:} Theorem \ref{main1} shows that $\bphi$ has
finite $H^{1,2}$ norm on each slice $[x_i,x_c)\ni x=$ constant.
Indeed we can write
\[ \int_\mathbb{R} \bphi^2(x,\rho)+\bphi_{,\rho}^2(x,\rho)\,d\rho \leq
C_3\hE(0)\] for some constant $C_3>0$ that depends only on
$G,\psi$ and $\ell$. Applying Lemma \ref{lemWald} yields
$|\bphi(x,\rho)|\leq \frac{C_3}{2}\hE(0)$,
$x\in[x_i,x_c),\rho\in\mathbb{R}$. As the bounding term is
independent of $x$, we can take the limit $x\to x_c$ and so extend
the bound to the Cauchy horizon $x=x_c$.\fin

To conclude this section, we point out that the results of Theorem
\ref{main1} and Theorem \ref{main2} can be extended to solutions
$\bphi$ that lie in $H^{1,2}(\mathbb{R})$ on each slice $x=$
constant. For these solutions, the field $\phi$ need not vanish at
the origin.

\begin{theorem}\label{main3}
Let $f\in H^{1,2}(\mathbb{R})$ and $g\in L^2(\mathbb{R})$. Then for
all $\kp\in[0,\min\{2,(1-x_c^2G^\prime(x_c))/2)\}]$ the initial
value problem consisting of (\ref{nweq}) with data
$\bphi|_{x=x_i}=f,\bphi_{,x}|_{x=x_i}=g$ has a unique solution
$\bphi \in C([x_i,x_c],H^{1,2}(\mathbb{R}))$. For such solutions,
the field values $\bphi(x,\rho)$ and energy $E[\bphi](x)$ satisfy a
priori bounds \begin{eqnarray*} \bphi(x,\rho)&\leq&
C_1\hE[\vec{\varphi}](0),\quad
(x,\rho)\in[x_i,x_c]\times\mathbb{R},\\ E[\bphi](x)&\leq& C_2
\hE[\vec{\varphi}](0),\quad x\in[x_i,x_c].\end{eqnarray*}
\end{theorem}

\noindent{\bf Proof:} The space $C^\infty_0(\mathbb{R})$ is dense
in both  $H^{1,2}(\mathbb{R})$ and $L^2(\mathbb{R})$ with the
appropriate norms (see Corollary 2.30 of \cite{adams} for the
former and Theorem 3.17 and Corollary 3.23 of \cite{adams} for the
latter result). Thus we can take a Cauchy sequence
$\{f_{(m)}\}_{m=0}^\infty$, (respectively
$\{g_{(m)}\}_{m=0}^\infty$) of functions in
$C^\infty_0(\mathbb{R})$ which converges in the $H^{1,2}$ norm to
$f$ (respectively, in the $L^2(\mathbb{R})$ norm to $g$). For each
$m\geq 0$, the functions $f_{(m)},g_{(m)}$ can be taken as initial
data for $\bphi,\bphi_{,x}$ respectively for the equation
(\ref{nweq}). Then the hypotheses of Corollary 2 and Theorems 3
and 4 above are satisfied, and we obtain a sequence of solutions
$\{\bphi_{(m)}\}_{m=0}^\infty$ with $\bphi_{(m)}\in
C^\infty([x_i,x_c)\times\mathbb{R},\mathbb{R})$ for all $m\geq 0$,
and such that each $\bphi_{(m)}$ has compact support on each $x=$
constant slice. Furthermore, the bounds of Theorem \ref{main1}
apply for each $m$: \be E[\bphi_{(m)}](x)\leq
C_1\hE[\vec{\varphi}_{(m)}](0),\quad
x\in[x_i,x_c].\label{seqapbnd}\ee Note that the constant $C_1$ is
the same for each $m$, and so by linearity
\[ E[\bphi_{(m)}-\bphi_{(n)}](x)\leq
C_1\hE[\vec{\varphi}_{(m)}-\vec{\varphi}_{(n)}](0),\quad
x\in[x_i,x_c],\] for all $m,n\geq 0$. The convergence properties
of $\{f_{(m)}\}_{m=0}^\infty$, and $\{g_{(m)}\}_{m=0}^\infty$
imply that the sequence of real numbers
$\{\hE[\vec{\varphi}_{(m)}](0)\}_{m=0}^\infty$ converges, and so
for each $x\in[x_i,x_c)$,
$\{(\bphi_{(m)},\bphi_{(m),x})\}_{m=0}^\infty$ is a Cauchy
sequence in the norm \[ \|(u,v)\|_E=\int_\mathbb{R} -\alpha v^2
+\gamma
u_{,\rho}^2+(\ell(\ell+1)e^\psi-(\kp^2-2\kp)\gamma+\kp\beta^\prime)u^2,\]
defined so that $\|(\bphi,\bphi_{,x})\|_E=E[\bphi](x)$. The space
$(\{(u,v):\mathbb{R}\to\mathbb{R}\times\mathbb{R}\},\|\cdot\|_E)$
is clearly complete, and so
$\{(\bphi_{(m)},\bphi_{(m),x})\}_{m=0}^\infty$ converges in this
space. This yields $\bphi(x)=\lim_{m\to\infty}\bphi_{(m)}(x) \in
H^{1,2}(\mathbb{R})$ for all $x\in[x_i,x_c]$. Furthermore, we can
take the limit $m\to\infty$ in (\ref{seqapbnd}) to obtain
\[ E[\bphi](x)\leq C_1\hE[\vec{\varphi}](0),\quad
x\in[x_i,x_c],\] which shows that
$\sup_{x\in[x_i,x_c]}\|\bphi(x)\|_{1,2}$ is bounded. We obtain the
pointwise bound by applying Theorem 4 to the sequence
$\{\bphi_{(m)}\}_{m=0}^\infty$ and taking the limit $m\to \infty$.
\fin

\begin{remarkthe}\label{spatials}
{\em The fact that the coefficients of (\ref{nweq}) are
independent of $\rho$ means that the spatial derivatives
$\partial_{\rho}^n\bphi$ also satisfy this equation, and so
Theorem 5 can be applied to these spatial derivatives. For
example, if we assume that $\bphi_{,\rho\rho\rho}$ and
$\bphi_{,x\rho\rho}$ are initially in $L^2$ (or more
appropriately, that initially $\bphi\in H^{3,2}$ and
$\bphi_{,x}\in H^{2,2}$), then we can apply Theorem 5 separately
to $\bphi, \bphi_{,\rho}$ and $\bphi_{,\rho\rho}$ to obtain
solutions for which $\bphi$ and its first two spatial derivatives
are bounded on $[x_i,x_c]\times\mathbb{R}$. Likewise, if we
specify data for (\ref{nweq}) with smooth compact support, then
the solution $\bphi$ and all its spatial derivatives will be
smooth, will have compact support on each slice $x=$ constant and
will be bounded on $[x_i,x_c]\times \mathbb{R}$. }
\end{remarkthe}

\section{Local energy measures}
In considering local observations of the energy content of the
multipole field $\phi$, two quantities are of relevance. These are
the flux of $\phi$ measured by an observer $O$ moving on a timelike
geodesic (with unit tangent $u^a$):
\[ \cf = u^a\partial_a\phi,\]
and the ``time-time" component of the energy-momentum tensor
$T_{ab}$ of the field. For the massless scalar field, we have
\[
T_{ab}=\partial_a\phi\partial_b\phi-\frac12g_{ab}g^{cd}\partial_c\phi\partial_d\phi,\]
and the energy measured by $O$ is
\[ \ce= u^au^bT_{ab}=(u^c\partial_c\phi)^2+\frac12g^{cd}\partial_c\phi\partial_d\phi.\]

Using the third part of Theorem 3 and an argument identical to
that used to prove Theorem \ref{main2}, we have global pointwise
bounds on $(x_c-x)^{1/2}\phi_{,x}$. From this and the line element
(\ref{lel2}), it is straightforward to show that the term
$g^{cd}\partial_c\phi\partial_d\phi$ is bounded on $\cmt$. Thus
finiteness of both $\ce$ and $\cf$ follows if and only if the term
$u^a\partial_a\phi$ is finite. In \cite{nolan-waters1}, we showed
that in the coordinates $x,\rho$, the components $u^a$ of the unit
vector field tangent to $O$ remain finite throughout $\cmt$ and
(crucially) at $\ch$. Thus $u^a\partial_a\phi$ is finite if and
only if both of the terms $\phi_{,x}$ and $\phi_{,\rho}$ are
finite. We know that the latter is finite, and so the local energy
($\ce$ or $\cf$) is finite at $\ch$ if and only if $\phi_{,x}$ is
finite at $\ch$.

We can show that this term must in fact be finite by rewriting the
wave equation (\ref{nweq}) as a first order transport equation for
$\bphi_{,x}$ with a source term involving only $\bphi$ and its
spatial derivatives. The fact that these terms are bounded enables
us to write down a formal solution of this transport equation and
hence demonstrate that $\bphi_{,x}$ is bounded in the limit as the
Cauchy horizon is approached. So we define $u:=\bphi_{,x}$. Then
(\ref{nweq}) reads
\be
\alpha u_{,x}+2\beta u_{,\rho}+(\alpha^\prime+2(1-\kp)\beta)u =
F(x,\rho),\label{trans}\ee where
\[ F(x,\rho) =
-\gamma\bphi_{,\rho\rho}-(\beta^\prime+2(1-\kp)\gamma)\bphi_{,\rho}+[\kp(2-\kp)\gamma+\kp\beta^\prime+\ell(\ell+1)e^\psi]\bphi.\]

The characteristics of this first order equation are defined by
\[ \frac{d\rho}{dx} = 2\frac{\beta}{\alpha},\]
giving
\[ \rho = \xi +\int_{x_i}^x 2\frac{\beta(s)}{\alpha(s)}\,ds = \xi
+ \lambda(x),\] where $\xi$ labels the different characteristics,
and gives the value of $\rho$ on a given characteristic as it
intersects the initial surface $x=x_i$. We note that these
characteristics are the outgoing radial null geodesics of the
spacetime. The equation (\ref{trans}) can then be written as an ODE
along individual characteristics: \[
\alpha\frac{d}{dx}\left\{u(x,\xi+\lambda(x))\right\}+(\alpha^\prime+2(1-\kp)\beta)u(x,\xi+\lambda(x))=F(x,\xi+\lambda(x)).\]
Defining
\[ J(x) = \exp \left\{ \int_{x_i}^x
\frac{\alpha^\prime+2(1-\kp)\beta}{\alpha}\,ds\right\},\] the
solution of the ODE can be written as follows: \be
J(x)u(x,\xi+\lambda(x))=u(x_i,\xi)+\int_{x_i}^x\frac{J(s)}{\alpha(s)}F(s,\xi+\lambda(s))\,ds.
\label{odesol}\ee

We now consider the limiting value of $u$ as the Cauchy horizon is
approached, using the solution (\ref{odesol}). We consider first
the case of smooth initial data with compact support. Noting that
$\lim_{x\to x_c}\lambda(x) = +\infty$, we see that approaching
$x=x_c$ at constant $\rho$ entails $\xi\to - \infty$. Thus for
values of $x$ sufficiently close to $x_c$, the characteristic
through the point $(x,\rho)$ meets the initial data surface
$x=x_i$ at an arbitrarily large, negative value of $\rho$,
yielding $u(x_i,\xi)=0$ (where we have appealed to the fact that
the data have compact support).

Next, we note that it follows from Remark \ref{spatials} and the
boundedness of the coefficients of $\bphi$ and its spatial
derivatives in $F$ that $F(x,\rho)$ is smooth, has compact support
on each surface $x=$constant and is bounded on
$[x_i,x_c]\times\mathbb{R}$. The bounding term will be of the form
$C_0\hE[\vec{\varphi}](0)+C_1\hE[\vec{\varphi}_{,\rho}](0)+C_2\hE[\vec{\varphi}_{,\rho\rho}](0)$,
where (as usual) the $C_i$ are constants depending only on the
functions $G,\psi$ and the angular mode number $\ell$. Here,
$\vec{\varphi}_{,\rho}$ is to $\bphi_{,\rho}$ as $\vec{\varphi}$
is to $\bphi$ (and likewise for the second derivative). Applying
the mean value theorem for integrals gives
\[ \int_{x_i}^x\frac{J(s)}{\alpha(s)}F(s,\xi+\lambda(s))\,ds =
F(x_*,\xi+\lambda(x_*))\int_{x_i}^x\frac{J(s)}{\alpha(s)}\,ds\]
for some $x_*\in[x_i,x]$.

From the definitions (\ref{aldef}) and (\ref{bedef}) of $\alpha$
and $\beta$ and the assumptions of Section 2, we have
\[ \frac{\alpha^\prime+2(1-\kp)\beta}{\alpha} =
\frac{\kp+x_c^2G^\prime(x_c)}{1+x_c^2G^\prime(x_c)}(x-x_c)^{-1}+O(1),\quad
x\to x_c.\] Thus
\[ J(x) =
J_0|x-x_c|^{\frac{\kp+x_c^2G^\prime(x_c)}{1+x_c^2G^\prime(x_c)}}+O(|x-x_c|^{\frac{\kp-1}{1+x_c^2G^\prime(x_c)}}),\quad
x\to x_c
\] where $0<J_0<\infty$ and so
\[ \lim_{x\to x_c} \frac{1}{J(x)}\int_{x_i}^x
\frac{J(s)}{\alpha(s)}\,ds = J_1,\] for some finite number $J_1$.

Combining the results of the last three paragraphs yields the
following: \be \lim_{x\to x_c}u(x,\rho) = F(x_{**},\rho)J_1 \ee
for some $x_{**}\in[x_i,x_c]$. This proves the following result.

\begin{theorem}\label{localenergy1}
Let $x_i\in(0,x_c)$ and let $f,g\in
C^\infty_0(\mathbb{R},\mathbb{R})$. Then the unique solution
$\bphi\in C^\infty(\mathbb{R}\times[x_i,x_c),\mathbb{R})$ of the
initial value problem consisting of (\ref{nweq}) with the initial
conditions $\bphi|_{x=x_i}=f,\bphi_{,x}|_{x=x_i}=g$ satisfies
$\lim_{x\to x_c} \bphi_{,x}(x,\rho) \in \mathbb{R}$ for all
$\rho\in\mathbb{R}$. Furthermore, this derivative satisfies an
{\em a priori} bound of the form \be \lim_{x\to x_c}
|\bphi_{,x}(x,\rho)|\leq
C_0\hE[\vec{\varphi}](0)+C_1\hE[\vec{\varphi}_{,\rho}](0)+C_2\hE[\vec{\varphi}_{,\rho\rho}](0),\quad
\rho\in\mathbb{R}.\label{localenergybd}\ee \fin
\end{theorem}

Again, invoking the density of $C^\infty_0(\mathbb{R})$ in certain
Sobolev spaces, we can obtain a more interesting result where the
field $\phi$ does not vanish at the origin.

\begin{theorem}
Let $x_i\in(0,x_c)$ and let $f\in H^{3,2}(\mathbb{R})$, $g\in
H^{2,2}(\mathbb{R})$. Then there is a unique solution $\bphi\in
C([x_i,x_c],H^{3,2}(\mathbb{R}))$ of the initial value problem
consisting of (\ref{nweq}) with the initial conditions
$\bphi|_{x=x_i}=f,\bphi_{,x}|_{x=x_i}=g$. This solution satisfies
\[ \lim_{x\to x_c} |\bphi_{,x}(x,\rho)|\leq
C_0\hE[\vec{\varphi}](0)+C_1\hE[\vec{\varphi}_{,\rho}](0)+C_2\hE[\vec{\varphi}_{,\rho\rho}](0),\quad
\rho\in\mathbb{R}.\]
\end{theorem}

\noindent{\bf Proof:} The existence and uniqueness part of the
result is an application of Theorem \ref{main3} and Remark
\ref{spatials}. We require the third derivative of $f$ to be in
$L^2$ to ensure finiteness of $\hE[\vec{\varphi}_{,\rho\rho}](0)$.
The {\em a priori} bound obtains by applying Theorem
\ref{localenergy1} to a sequence of solutions
$\{\bphi_{(m)}\}_{m=0}^\infty$ of (\ref{nweq}) which satisfy
$\bphi_{(m)}(x_i,\rho)\to_{m\to\infty}f$ in $H^{3,2}(\mathbb{R})$
and $\bphi_{(m),x}(x_i,\rho)\to_{m\to\infty}g$ in
$H^{2,2}(\mathbb{R})$. The bound (\ref{localenergybd}) applies to
each member of this sequence, and so applies to the solution
$\bphi$ in the limit. \fin

\section{Comment on massive fields and non-minimal coupling}
We consider briefly a massive minimally coupled field $\phi$,
satisfying the Klein-Gordon equation $\Box\phi-m^2\phi=0$. The
mass parameter $m>0$ introduces a length scale, and so must break
self-similarity. This is reflected in the presence of a radially
dependent term in the wave equation: the massive field $\phi$
satisfies an equation differing from (\ref{weq}) only in the term
in $\phi$:
\be
\alpha\phi_{,xx}+\cdots-(m^2e^{2\rho}+\ell(\ell+1))e^\psi\phi=0.
\label{masseq}\ee The presence of this additional term invalidates
some but not all of the results above. We are able to retain the
results relating to wave packets, i.e.\ results relating to the
case where the initial data are smooth with compact support. In
this case, Theorem \ref{corrEUforbphi} applies. An amended version
of Corollary \ref{corrEhat} also applies: we can obtain a bound of
the form
\[ \hE(t)\leq C_0e^{B_0t}\mu(\vec{\varphi}(0))\hE(0),\]
where $\mu(\vec{\varphi}(0))$ is the Lebesgue measure of the
support of the initial data $\vec{\varphi}(0)$. This number will
determine - via the characteristics of the system (\ref{system}) -
a maximum for $\rho$ on the support of the solution
$\vec{\varphi}$ at time $t$, and so will determine a bound for
$|B(t)|$ that will depend on $t$ and $\mu(\vec{\varphi}(0))$. This
will have the form indicated above. Noting that Lemma
\ref{lemEgrow} remains valid for $m\neq 0$, we can conclude that
Theorems \ref{main1} and \ref{main2} hold for a massive, minimally
coupled scalar field. However the arguments used to prove Theorem
5 break down due to the presence of the Lebesgue measure of the
support of the initial data in the bound above. It would be of
interest to determine if another approach could be used to obtain
the equivalent of these results in the massive case. This should
be feasible, especially as what one is most interested in doing is
extending the function space at the origin $\rho=-\infty$, where
the radially dependent term is exponentially small. The argument
of Theorem 6 also breaks down, as we will not have a global bound
for the function corresponding to $F$.

We could also consider other couplings, so that $\phi$ satisfies
e.g.\ $\Box\phi-\xi R\phi=0$, where $\xi$ is a constant and $R$ is
the Ricci scalar. Here, an additional term of the form $q(x)\phi$
is present in the wave equation, so self-similarity is preserved
(as expected). This will make a crucial difference in Lemma
\ref{lemEgrow}, and it does not seem possible to draw any general
conclusions as to the continued applicability of this lemma, and
hence of the validity any of our principal results. We expect that
similar results could however be obtained on a case by case basis,
where the background geometry is specified.

\section{Conclusions}

We have shown that in self-similar collapse to a naked singularity,
the multipoles of a massless scalar field propagating on the
background spacetime remain finite as it impinges on the Cauchy
horizon. This has been shown to be true of different measures of the
multipole field: its point-values, its $L^2$ norm and different
local ($\ce,\cf$) and global ($E$) measures of its energy content.
We have concentrated on the field $\bphi = \phi e^{\kp\rho}$, and so
it is instructive to consider the implications for $\phi$ of (in
particular) the pointwise bounds obtained. We have shown that
$r^\kp\phi(x,r)$ and $r^\kp\phi_{,x}(x,r)$ are bounded on the Cauchy
horizon for all $\kp\in[0,\kp_*)$ where $\kp_*>0$ is determined by
the background geometry. Thus the values of $\phi$ and $\phi_{,x}$
on the Cauchy horizon can only diverge at the singularity. In
particular, any possible such blow up does not make itself felt
along the Cauchy horizon: it is confined to the central singularity.
Similar (but more limited) results apply also for massive scalar
fields. The question of whether or not these results will also apply
to the full field obtained by resumming the multipoles remains open,
but it should be noted that the multipoles themselves have an
independent physical significance. For example, one would expect
that the scalar {\em radiation} field is dominated by the $\ell=0$
contribution.

We interpret our results as providing evidence that these naked
singularities may be stable under perturbations of the background
spacetime, and so may constitute a challenge to the cosmic
censorship conjecture. In particular, we consider that these results
add weight to the existing evidence that GRLP spacetimes admit
stable naked singularities \cite{harada}. As there are no unstable
modes of perturbations impinging on the singularity, it is likely
that these spacetimes give rise to the type of initial data studied
here. This is also possible for the single-unstable mode critical
spacetimes \cite{carsten}. In these spacetimes, the single unstable
mode shows a characteristic divergence as $r\to 0$ along surfaces
$x=$ constant. If this divergence is sufficiently mild, this mode
would correspond to initial data of the type studied here, where
weighting by a power of $r$ brings the data into a certain Sobolev
space. The situation described in the previous paragraph could then
hold, where the divergence at the singularity is not felt along the
Cauchy horizon, yielding stability of the horizon.

Let us suppose then that the results found here for a minimally
coupled scalar field carry over to the metric and matter
perturbations of the background spacetime. This would indicate only
{\em linear} stability of the naked singularity. An entirely
different picture may emerge when full (non-linear) stability is
considered. For example, in the case where the background energy
momentum tensor is that of a minimally coupled massless scalar field
(Einstein-Klein-Gordon spacetime), examples of naked singularities
are present in the self-similar case \cite{christo1,brady2}. Our
results apply to these examples, and so provide evidence of linear
stability. However these have been proven to be non-linear unstable
\cite{christo2}. Thus our analysis is probably best interpreted as
demonstrating the validity of a {\em necessary} - but nonetheless
nontrivial - condition for stability.

Finally, we note the connection with generalised hyperbolicity
\cite{clarke}/wave regularity \cite{ishibashi-hosoya,
stalker}/quantum regularity
\cite{wald3,horowitz-marolf,ishibashi-wald} of naked singularities.
These are broadly similar ways of characterising whether or not a
naked singularity does in fact provide a fundamental barrier to
predictability in spacetime. As a next step in the study of naked
singularities in self-similar spacetimes, it would be of interest to
determine whether or not these singularities pass the tests of
generalised hyperbolicity/wave regularity/quantum regularity. A
minimum requirement for passing such tests is that the field $\phi$
remains intact as it impinges on the Cauchy horizon. As shown above,
this requirement holds. The more complicated question of the
well-posedness of the field to the future of the Cauchy horizon
remains to be answered.

\ack I thank Mihalis Dafermos for comments on the interpretation
of the results above, Hans Ringstr\"om for suggesting the use of a
norm of the general form $E$, and John Stalker for useful
conversations.

\end{document}